\documentclass[preprint,final,5p,times,twocolumn]{elsarticle}

\usepackage[colorlinks,linkcolor=blue,anchorcolor=blue,citecolor=blue,urlcolor=blue]{hyperref}
\usepackage{graphicx}
\usepackage{epstopdf}
\usepackage{dcolumn}
\usepackage{textcomp}
\usepackage{gensymb}
\usepackage{booktabs}
\usepackage{amsmath}
\usepackage{amssymb}
\usepackage{color}
\usepackage{hyperref}
\usepackage{url}
\usepackage{siunitx}
\usepackage{subfigure}
\usepackage{longtable}
\usepackage{array}
\usepackage{booktabs}
\usepackage{verbatim}
\usepackage{threeparttable}
\usepackage{setspace}

\emergencystretch=\maxdimen
\usepackage{dcolumn}
\newcolumntype{x}{D{.}{.}{6.6}}
\newcolumntype{y}{D{.}{.}{4.5}}
\newcolumntype{z}{D{.}{.}{5.7}}
\newcolumntype{f}{D{.}{.}{7.9}}
\newcolumntype{e}{D{.}{.}{5.6}}

\journal{Nuclear Instruments and Methods in Physics Research Section A}

\begin{document}

\begin{frontmatter}

\title{Construction and commissioning of the collinear laser spectroscopy system at BRIF}

\author[PKU]{S.~J.~Wang}
\author[PKU]{X.~F.~Yang\corref{cor1}}\ead{xiaofei.yang@pku.edu.cn}
\author[PKU]{S.~W.~Bai}
\author[PKU]{Y.~C.~Liu}
\author[PKU]{P.~Zhang}
\author[PKU]{Y.~S.~Liu}
\author[PKU]{H.~R.~Hu}
\author[CIAE]{H.~W.~Li}
\author[CIAE]{B.~Tang}
\author[CIAE]{B.Q.~Cui}
\author[CIAE]{C.~Y.~He}
\author[CIAE]{X.~Ma}
\author[PKU]{Q.~T.~Li}
\author[PKU]{J.H.~Chen}
\author[PKU]{K.~Ma}
\author[PKU]{L.S.~Yang}
\author[PKU]{Z.Y.~Hu}
\author[PKU]{W.L.~Pu}
\author[PKU]{Y.~Chen}
\author[PKU]{Y.~F.~Guo}
\author[PKU]{Z.~Y.~Du}
\author[PKU]{Z.~Yan}
\author[CIAE]{F.L.~Liu}
\author[CIAE]{H.R.~Wang}
\author[CIAE]{G.Q.~Yang}
\author[PKU]{Y.~L.~Ye}
\author[CIAE]{B.~Guo}

\cortext[cor1]{Corresponding author}
\address[PKU]{School of Physics and State Key Laboratory of Nuclear Physics and Technology, Peking University, Beijing 100871, China}
\address[CIAE]{China Institute of Atomic Energy (CIAE), P.O. Box 275(10), Beijing 102413, China}

\begin{abstract}
We have constructed a collinear laser spectroscopy (CLS) system installed at the Beijing Radioactive Ion-beam Facility (BRIF), aiming to investigate the nuclear properties of unstable nuclei. The first on-line commissioning experiment of this system was performed using the continuous stable ($^{39}$K) and unstable ($^{38}$K) ion beams produced by impinging a 100-MeV proton beam on a CaO target. Hyperfine structure spectra of these two isotopes are reasonably reproduced, and the extracted magnetic dipole hyperfine parameters and isotope shift agree with the literature values. The on-line experiment demonstrates the overall functioning of this CLS system, opening new opportunities for laser spectroscopy measurement of unstable isotopes at BRIF and other radioactive ion beam facilities in China.
\end{abstract}


\end{frontmatter}

\section{Introduction}
Nuclear spins, magnetic dipole moments, electric quadrupole moments and mean-square charge radii, are fundamental properties of atomic nuclei. These experimental observables are essential for understanding nuclear structure and nucleon-nucleon interactions~\cite{Neyens2003,JPG2010,JPG2017} and for stringent tests of state-of-art nuclear theories~\cite{Ca-radii2016, K-radii2021,Cu-radii2020}. Collinear laser spectroscopy (CLS) is a well-established tool to measure these nuclear properties simultaneously in a nuclear-model independent way, by probing the hyperfine structure (HFS) and isotope shift arising from the interaction between nucleus and surrounding electrons. Since the first demonstration of CLS in 1970s~\cite{CLS1,CLS2}, it has been extensively applied to investigate the structure of short-lived nuclei far from the valley of $\beta$-stability~\cite{PPNP2016}. 

CLS experiments rely on the measurement of Doppler-free optical spectra by overlapping a low-energy (few tens of keV) atom or ion beam with a high-resolution laser beam in a collinear (or anti-collinear) geometry~\cite{JPG2017,BECOLA,CRIS,IGISOL}. Such studies have been extensively carried out at radioactive ion (RI) beam facilitates based on the Isotope Separation On-Line (ISOL) technique, such as CERN-ISOLDE~\cite{ISOLDE2017}, IGISOL of JYFL~\cite{IGISOL2014} and ISAC of TRIUMF~\cite{ISAC-TRIUMF2014}, where the RI beams are delivered with high quality (small emittances) at a suitable energy (e.g. $\sim$30-60~keV). The production of bunched ion beams using RFQ cooler-bunchers~\cite{ISCOOL} has dramatically enhanced the experimental sensitivity of the CLS technique~\cite{JPG2010}. Furthermore, the installation of the sophisticated techniques (e.g. gas catcher) for converting the high-energy RI beams into low-energy beams, has enabled the implementation of CLS technique into PF (projectile fragmentation)-type RIB facilities~\cite{BECOLA}, which produce the in-flight RI beam at tens or hundreds of MeV per nucleon. In recent years, the combination of the high-resolution CLS with the high-sensitivity resonance ionization spectroscopy (RIS) has further broadened the applicability of the CLS for more exotic nuclei~\cite{CRIS}. The relevant CLS setups have been widely established and planned on the existing and the on-going next generation RI beam facilities worldwide~\cite{JPG2017,BECOLA,CRIS,IGISOL,CLS-RIKEN,CLS-RAON,BISOL,LaSpec}.

\begin{figure*}[!t]
\includegraphics[width=0.99\hsize]{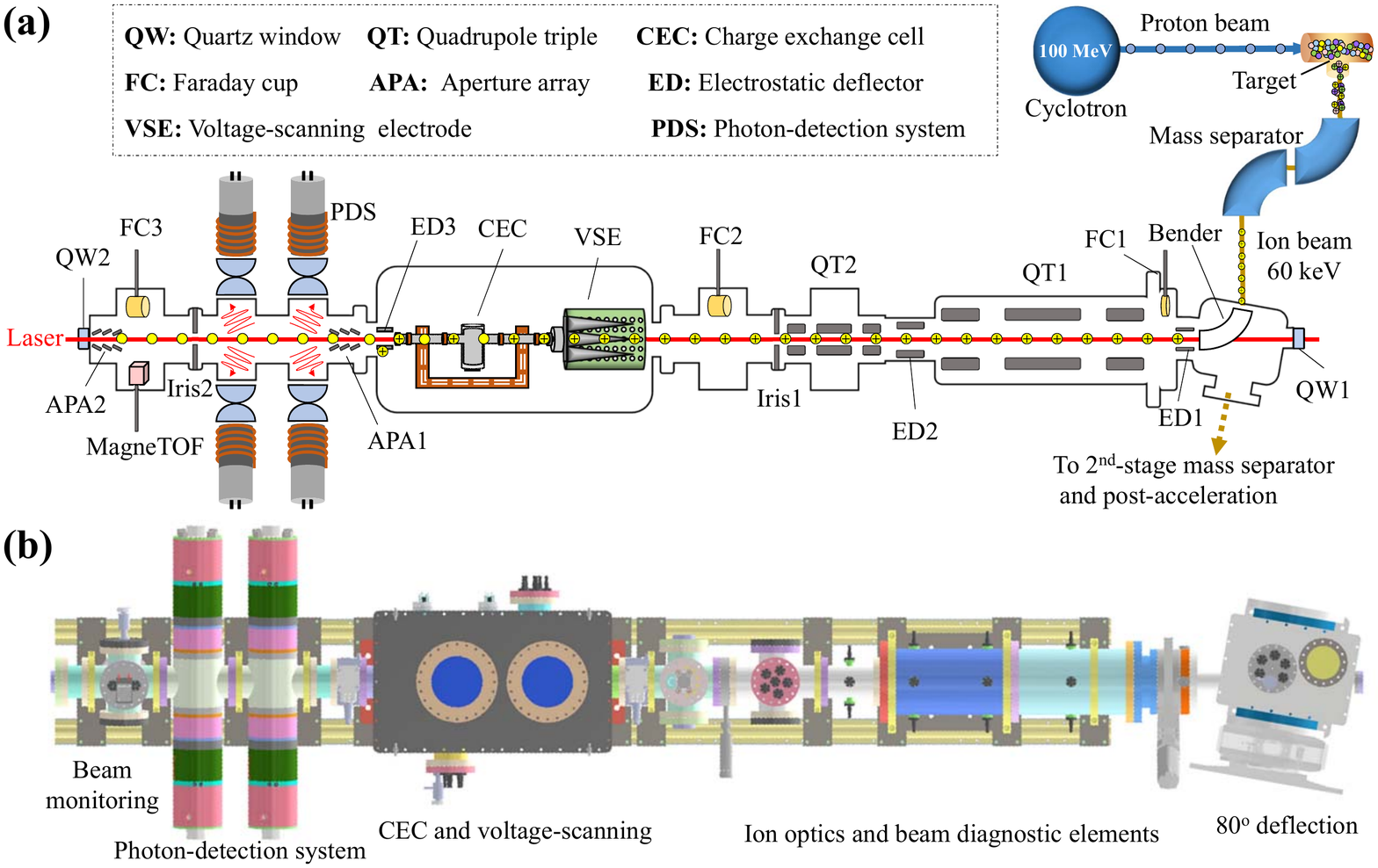}
\caption{(a): Schematic top view of the CLS system commissioned at BRIF. The RI isotopes are produced by impinging the 100-MeV proton beam on a solid target, which are subsequently surface ionized, extracted and accelerated to 60~keV. After mass separation, the ion beam is delivered into the CLS system, where it overlaps in space with a laser beam in an anti-collinear configuration. In the CLS beamline, the velocity of the ion beam can be tuned by a voltage-scanning electrode before injecting into the charge-exchange cell, where the ions are neutralized via collision with alkali-vapor. The neutral atoms are then resonantly exited by the laser in the photon-detection region. The fluorescent photons emitted from the excited atoms are collected by the photon detection system. See text for more details. (b): CAD drawing of the CLS beamline.}
\label{Fig1}
\end{figure*}

Research on unstable nuclei in China has greatly benefited from RI beam facilities, the Heavy Ion Research Facility in Lanzhou (HIRFL) using the PF technique~\cite{HIRFL,PRL-RIBILL-2020} and the Beijing Radioactive Ion-beam Facility (BRIF) using the ISOL technique~\cite{NIMB-BRIF-2016,NIMB-BRIF-2020}. Meanwhile, two next-generation RI beam facilities, i.e. HIAF (High Intensity heavy ion Accelerator Facility)~\cite{HIAF} and BISOL (Beijing Isotope-Separation-On-Line neutron-rich beam facility)~\cite{BISOL} are currently under construction and planned, respectively, which will offer new opportunities for the exploration of unstable nuclei. As laser spectroscopy apparatus has not yet been established at the RI beam facilities in China, we thus have developed a new CLS system at BRIF, part of which was already tested with the bunched stable ion beams provided by an offline laser ablation ion source, as reported recently in Ref.~\cite{CLS-PKU}.

The BRIF facility has successfully produced the RI beams of $^{20-22}$Na and $^{36-38}$K isotopes by impinging a 100-MeV proton beam on MgO and CaO targets, respectively~\cite{BRIF1, BRIF2}. Implementation of the UC$_{\rm x}$ target into the BRIF is on-going, which has the potential to produce a wide range of RI isotopes in various mass regions. BRIF is thus an ideal venue for on-line laser spectroscopy experiment at the first stage. In the present work, the newly developed CLS system was installed at BRIF facility, and the first on-line commission experiment was successfully performed with continuous stable ($^{39}$K) and unstable ($^{38}$K) ion beams from BRIF. The HFS of $^{38,39}$K were reasonably well reproduced, and the extracted HFS parameters and isotope shift between $^{38}$K and $^{39}$K are in line with literature~\cite{K-radii2021,K-AB-2019}.

\section{Experimental setup}
A layout of the whole experimental setup and a schematic drawing of the CLS system at BRIF are presented in Fig.~\ref{Fig1}(a) and (b), respectively. The CLS system is designed using ConFlat (CF) standard for high vacuum, reaching $\sim$10$^{-8}$~mbar in the offline test~\cite{CLS-PKU}. Achieving high vacuum is a prerequisite for the planned upgrade of the system to achieve high-sensitivity collinear resonance ionization spectroscopy in the near future~\cite{CRIS}.

As shown in Fig.~\ref{Fig1}, a 100-MeV proton beam impinges upon a solid target, producing various reaction products. By using a surface ion source, the produced isotopes can be ionized, extracted and accelerated up to energies of 150 keV. The accelerated ions in continuous mode are then mass separated using the isotope separator and delivered into the CLS setup, where they can be partially neutralized in-flight by passing through a charge-exchange cell (CEC) filled with alkali (Na or K) vapor. A continuous-wave high-resolution laser beam locked to a particular frequency is overlapped with the ion (atom) beam. Prior to neutralization, the velocity of the beam can be tuned by applying a variable potential to the voltage-scanning electrode upstream of the CEC. This is to tune the Doppler-shifted laser frequency to match the transition frequency in the reference frame of the isotope of interest. The emitted fluorescent photons from the resonantly excited atoms are detected by the photon detection system. Photon signals are collected and processed by the Data Acquisition system (DAQ) with respect to the tuning voltage to obtain the HFS spectrum for the studied isotope.

\subsection{BRIF}
BRIF, the Beijing Radioactive Ion-beam Facility, is the first ISOL-type RI beam facility in China, which was constructed by the China Institute of Atomic Energy~\cite{NIMB-BRIF-2016,NIMB-BRIF-2020}. It consists of a 100-MeV high-intensity (up to 200~$\mu$A) proton cyclotron (referred to as CYCIAE-100), a target station with surface ion source, a first-stage mass separator with mass-resolving power of few thousands, and a second-stage mass separator with 20000 mass-resolving power. The high-resolution mass separator is followed by a pre-existing 13-MV tandem accelerator to enable delivery of post-accelerated RI beam~\cite{BRIF2}. BRIF was first commissioned in 2015 by producing the unstable $^{20-22}$Na and $^{36-38}$K isotopes. This led to a successful day-one experiment to study the exotic decay mode of the neutron-deficient $^{20}$Na isotope~\cite{BRIF1}. 

\subsection{CLS beamline: ion-optics and beam diagnostics}\label{sec:optics}
As shown in Fig.~\ref{Fig1}, the CLS setup is installed down stream of the first-stage mass separator. It is about 4-m long, consisting of multiple vacuum chambers with ion optics and ion-beam diagnostic elements, a charge-exchange chamber (CEC and voltage-scanning electrode) and a photon-detection system.

\begin{figure}[!b]
\center
\includegraphics[width=0.5\textwidth]{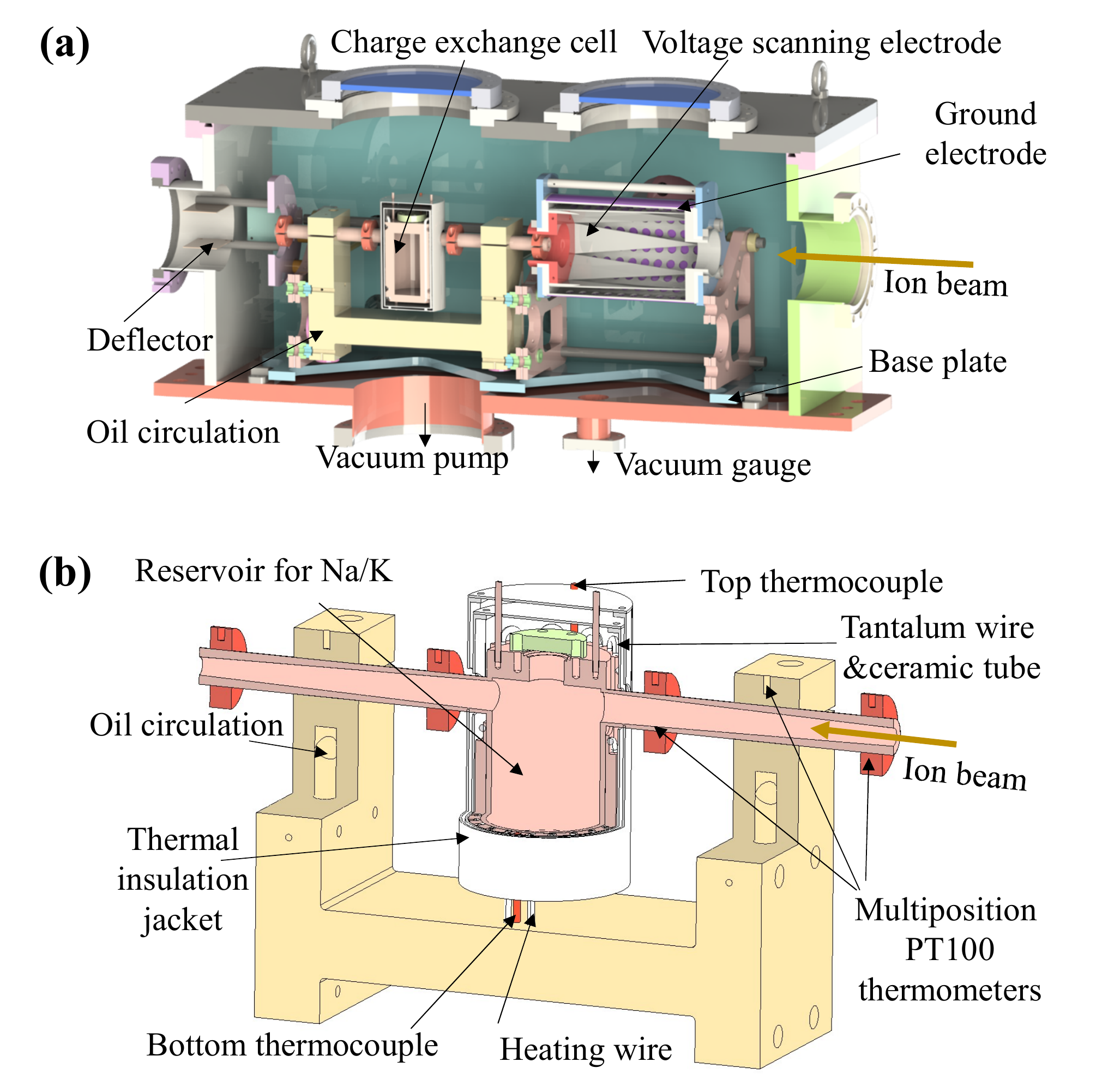}
\caption{(a): Overview of the charge-exchange chamber, consisted of voltage-scanning electrode, CEC and deflector electrode. (b): 3/4 view of the CEC with all the details included. See text for details.}
\label{Fig2}
\end{figure}

The incoming continuous ion beam from BRIF is guided to the CLS beamline by a pair of parallel 80$^\circ$ electrostatic bender plates. The spacing between the plates is 26~mm, and the maximum potential applied to each plate is 10~kV supplied by $\pm$ 10~kV HV modules (EHS 40100p/n, iseg). The CLS beamline is a new device inserted into the initial layout of the BRIF system (Fig.~8 in Ref.~\cite{NIMB-BRIF-2016}). Thus, there is only a very limited space ($\sim$39 cm long as shown in Fig.~\ref{Fig1}(b)) to embed the deflection chamber. In addition, the bender plates in the chamber are required to be removable in order to alternatively deliver the ion beam to the second-stage mass separator for the post-acceleration. The plates are therefore mounted onto a linear actuator to enable them to be installed and removed in a reproducible manner. A pair of electrostatic deflectors (ED1) follows to correct the ion beam direction after the 80$^\circ$ bend. The ion beam is then transported through the CLS beamline using a series of electrostatic elements. Two sets of quadrupole triplets (QT1 and QT2) lens are used to control the ion beam focus, and the $x-y$ electrostatic deflectors (ED2) provide minor adjustment of the ion beam trajectory. The potential applied to these QT and ED electrodes can be up to 6~kV provided by $\pm$~6~kV HV modules (HTP-EHS F0-60n/p, iseg). All the 10~kV and 6~kV HV modules are controlled by a HV crate (ECH238, iseg).

Ion beam diagnostic systems play a crucial role in the setup and include three Faraday cups (FC), two iris diaphragms and one ion detector (14924 MagneTOF Mini, ETP). Due to space limitation, FC1 is made of a thin copper plate and installed inside of a double-faced flange, while FC2 and FC3 are equipped with secondary-electron suppressor. The ion beam current is recorded by a picoammeter (6485, Keithley), which, together with the voltage controller for ion-optics elements, can be operated by a Python-based program. Two iris diaphragms with tunable diameter up to 20~mm function as the collimators for the alignment of the ion and laser beams, and can also used to check the beam diameter.

At the end of the CLS beamline, a MagneTOF ion detector is mounted at the bottom of a six-way cross chamber. This detector can be used to assess the neutralization efficiency of CEC. In addition, the ion detector is found to be sensitive to the variation in beam intensity, even when it is removed from the ion beam path. This is due to the capability of the MagneTOF to detect the scattered particles arisen from the ion beam hitting the materials along the beamline. Thus, the ion detection can act as an on-line beam intensity monitor during experiments. This function was found to be essential for the success of the present on-line commissioning measurement.

\begin{figure}[!t]
\center
\includegraphics[width=0.5\textwidth]{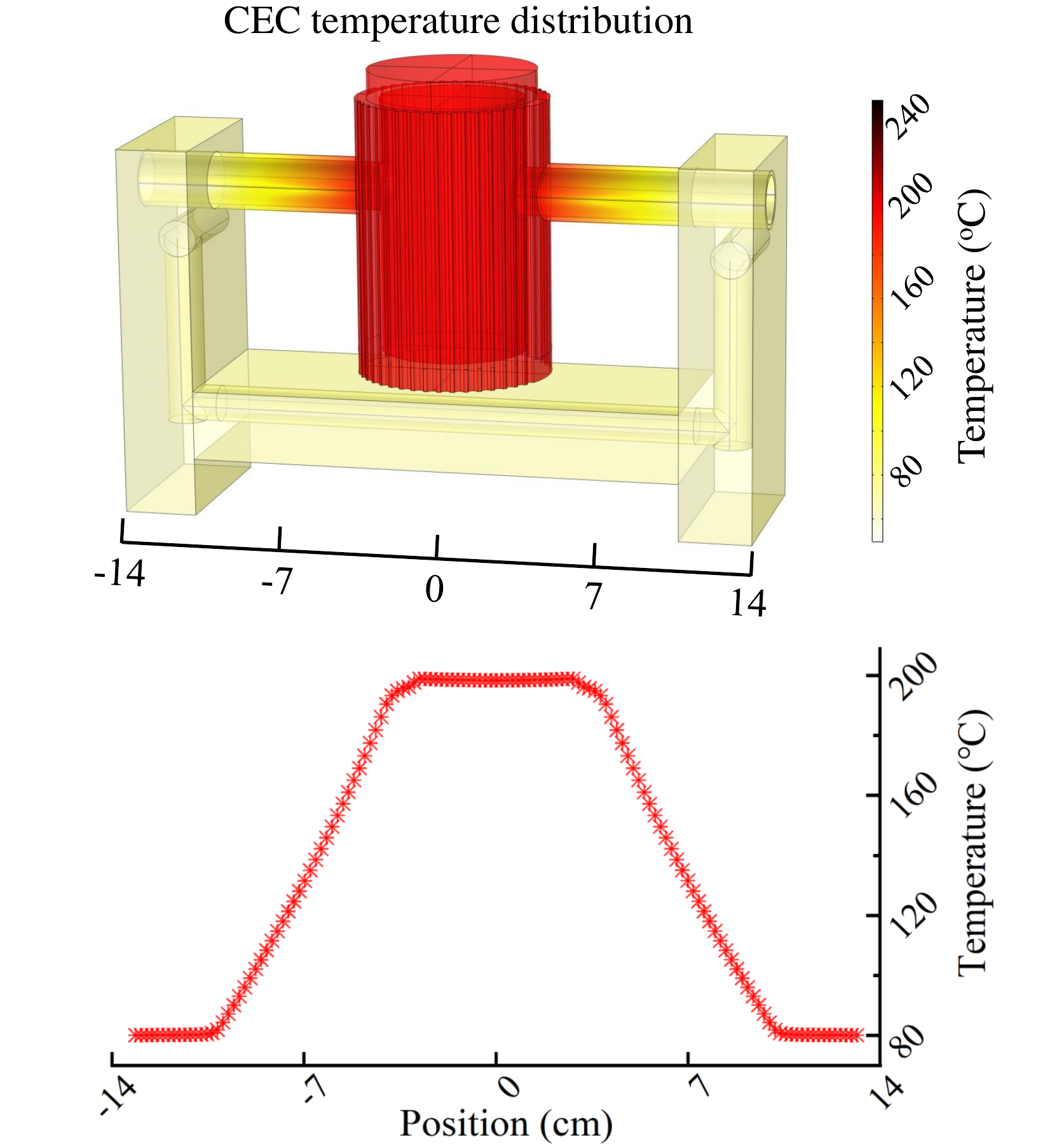}
\caption{Temperature distribution of the CEC based on simulations with COMSOL Multiphysics~\cite{COMSOL}. The isothermal contours (top panel) and temperature curve (bottom panel) are obtained for a central heating temperature of 200$^\circ$C and an oil circulation of 80$^\circ$C at the two ends of the CEC.}
\label{Fig3}
\end{figure}

\subsection{Voltage-scanning system and charge-exchange cell}\label{sec:cec}

Suitable transitions for laser spectroscopy often exist in the neutral atomic form of a given element. Therefore, for an experiment probing the atomic transition, the incoming ions need to be neutralized prior to laser excitation. Most of CLS experiments are performed by varying the ion-beam velocity to tune the Doppler-shifted laser frequency. This is realized by applying a scanning voltage to an acceleration or deceleration electrode (voltage-scanning electrode).

For this purpose, a charge-exchange chamber is installed just before the photon-detection region, as indicated in Fig.~\ref{Fig1} and presented in Fig.~\ref{Fig2}(a). Along the ion beam direction, voltage-scanning electrode, CEC and deflector electrode are mounted in sequence. The voltage-scanning electrode and the CEC are at an equal potential and isolated from the base plate at ground.\\

\subsubsection{Voltage-scanning electrode}
The voltage-scanning electrode installed upstream of the CEC \mbox{(Fig.~\ref{Fig2}(a))} is a slightly modified version of the operational system at VITO-ISOLDE~\cite{NIMA-GINS-2019}. It is a crown-shaped electrode, consisting of eight triangular spikes and an octagonal mounting base to define the equipotential electrical field. A cylinder grounding electrode, covering the whole voltage-scanning electrode, is isolated from this biased electrode with a ceramic insulator. This design is different from the commonly used one, i.e. a series of ring electrodes connected through a resistor chain~\cite{BECOLA,ALTO,JPG2017}. This is based mainly on the requirement of not changing the trajectory of ion beam along a long distance during the voltage scanning (velocity tuning of the ion beam), as has been demonstrated by simulation and  experimental tests in Ref.~\cite{NIMA-GINS-2019}.

\subsubsection{Change exchange cell}

It is known that heating the CEC might largely affect the vacuum condition, which is an intractable problem for experiments under high vacuum. Therefore, special care has been taken for the design of the CEC. Figure~\ref{Fig2}(b) depicts the details of CEC components. The operational temperature of this CEC can be up to 400$^{\circ}$C. In the center, there is a cylindrical reservoir made of stainless steel to store sodium or potassium. This reservoir is installed in a coaxial outer shell, consisting of thirty evenly spaced semicircular and vertical grooves. Ceramic tubes are fixed in these grooves and a tantalum heating wire is vertically wound through these ceramic tubes around the shell. This layout ensures electric isolation from the high potential of the CEC and uniform heating across the center of the cell. Two thermocouples are mounted on the top and bottom of the shell to monitor the temperature in the center of the CEC. The reservoir and the shell are fully covered by a heating insulation jacket in order to minimise the effects on the high vacuum during CEC heating. To prevent the sodium or potassium vapor from diffusing into the rest of the beamline, a cooling system is connected to the two ends of the CEC, which circulates heat transfer liquid inside of a copper heatsink using a thermostat circulator. In this way, a lower temperature can be kept at the two ends of the CEC. This end-temperature should be just above the melting point of potassium (63.2~$^{\circ}$C) or sodium (97.8~$^{\circ}$C), such that alkali vapor diffusing towards the ends of the CEC will condense into a liquid and flow back to the center of the CEC along the axially wedged inner wall of CEC tube. Multiple thermometers (PT100) are positioned at different locations of the CEC, which allows the temperature distribution of the CEC to be continuously monitored. Two collimators with 8~mm diameter are mounted at both ends of the CEC to further help prevent diffusion of alkali vapor into the rest of the vacuum chamber. Based on all these implementations, the diffusion of alkali vapor can be minimized and at the same time high vapor density could be maintained in the middle part of the CEC.

In order to gain more knowledge about the CEC, simulations were performed beforehand based on the software of COMSOL Multiphysics~\cite{COMSOL}. Figure~\ref{Fig3} illustrates the isothermal contours and the temperature distribution on the surface of the CEC under the condition: a central heating temperature at 200$^{\circ}$C and an end temperature at 80$^{\circ}$C. Note that the temperature variation is basically controlled by the heat-conductive metal structure and thus the potassium vapour is not included in the simulation for this qualitative illustration. As expected, within the central area (approximately 80~mm) of the CEC, a uniform temperature distribution on the surface of the CEC is achieved, ensuring a high-density interaction zone for ion neutralization. The temperature drops towards the ends of CEC and is stabilized at 80~$^{\circ}$C. The operation and the performance test of CEC and voltage scanning system will be further detailed in the following section~\ref{cectest}. 

\begin{figure}[t!]
\center
\includegraphics[width=0.5\textwidth]{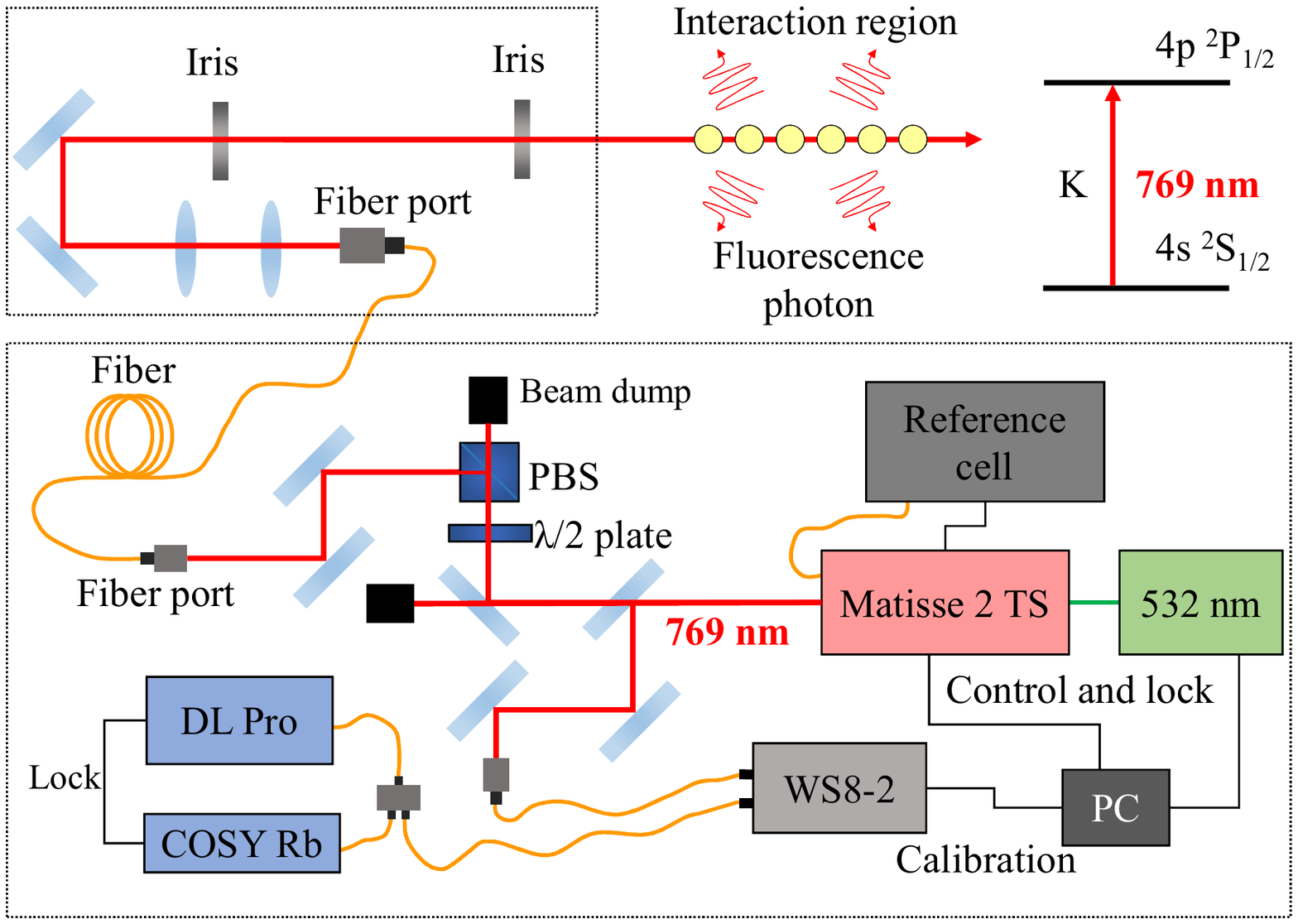}
\caption{Layout of the laser system used for the on-line experiment at BRIF. See text for details.}
\label{Fig4}
\end{figure}

\subsection{Laser system}\label{sec:laser}
The laser setup used for this on-line experiment is schematically shown in \mbox{Fig.~\ref{Fig4}}. Pumped by a 532-nm laser (20~W, Millenia Prime, Spectra Physics), the continuous-wave (cw) titanium-sapphire (Ti:Sa) laser (Matisse 2 TS, Sirah Lasertechnik) can generate the single-mode fundamental laser light within the wavelength range of 650-1020~nm by using different mirror sets. As introduced in Ref.~\cite{CLS-PKU}, further frequency-doubling of this fundamental light can be achieved by using the Wavetrain 2 (Sirah Lasertechnik) cavity, offering a laser light with 325-510~nm wavelength. The short-term stability of the Ti:Sa cavity is achieved by using an external reference cell. Additionally, a high-precision wavelength meter (WS8-2, HighFinesse) is adopted for the long-term stabilization of the laser. In order to compensate for drifts in the wavelength meter resulting from temperature and pressure fluctuations, a commercial saturation absorption spectroscopy unit, including a tunable diode laser (DLPRO780, TOPTICA Photonics AG), a temperature-controlled vapor cell (COSY, TOPTICA Photonics AG), is used~\cite{CLS-PKU,K-AB-2019}. The polarization of the laser beam is selected by using a combination of a half-wave plate and polarizing beam-splitter which also serves as a laser power attenuator. Due to space constraint, a five-meter-long optical fiber is used to transport the laser light to the optical breadboard mounted at the end of the CLS beamline. A series of the optical elements (mirrors, lens, iris diaphragms) are used to adjust the laser position and focus through the CLS beamline. An aperture is mounted at the entrance window (QW2) at the end of the CLS beamline, to further aid laser alignment. The laser power at the exit window (QW1) of the beamline (Fig.~\ref{Fig1}) can be monitored in real time using a laser power meter (VEGA, OPHIR). 

\begin{figure*}[t!]
\center
\includegraphics[width=0.95\hsize]{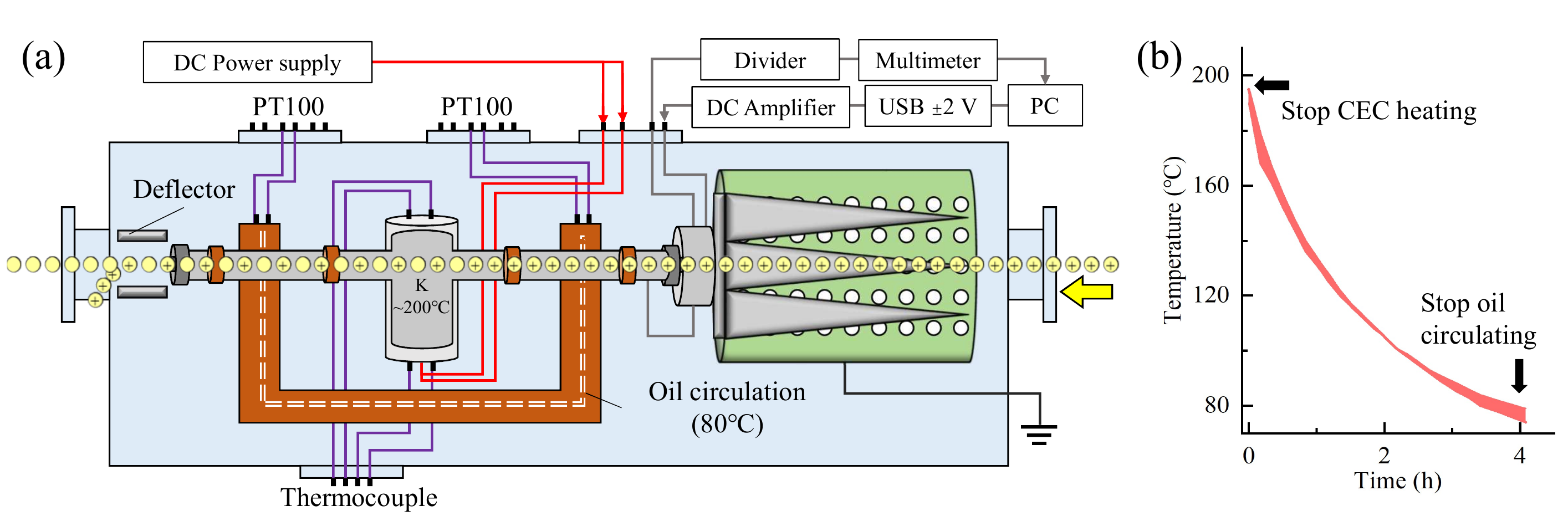}
\caption{(a) Schematic diagram of the voltage-scanning electrode and CEC, and of the heating system for the CEC. The voltage-scanning electrode and CEC is floating at the same potential between $-$2~kV and $+$2~kV supplied by the DC amplifier, which is controlled by the DAQ program via the USB device. A DC power supply is used to heat the CEC through the tantalum wire. Temperatures at different location of the CEC is monitored using two types of thermometers, PT100 and thermocouple. Ends of these themometers attached to the CEC is sealed with ceramics in order to be isolated from the ground. (b) The descent curve of the temperature read from the CEC center after stopping the CEC heating. The band width gives the difference between the read-values from two thermocouples at the top and bottom of the CEC.}
\label{Fig5}
\end{figure*}

\subsection{Photon-detection system and DAQ system}\label{sec:daq}
The photon detection system in the on-line CLS setup consists of four detection units. Details of the design, composition and operation of this detection unit have been reported in Ref.~\cite{CLS-PKU}. In brief, each unit consists of two aspheric lenses (AL100100, N-BK7, Thorlabs), a PMT (R943-02, Hamamatsu) assembled with a socket (E2762-506, Hamamatsu). For the wavelength of 769~nm used in the measurements presented here \mbox{(Sec.~\ref{sec:test})}, the quantum efficiency of the PMT is just above 10\%. To reduce dark counts, the PMTs are cooled to a low temperature of $-28^\circ$C with ethanol circulated by a cooling thermostat (ECO RE 630S, LAUDA). Note that, for the on-line detection system, the closest distance between the position along the beam direction and the first lens was shortened to enlarge the geometrical coverage. As a results, a total geometrical efficiency of the four detection units is simulated to be larger than 40\%. The distance between the lens and the PMT was adjusted by changing the thickness of the spacer to be 49.3~mm for optimal detection of 769-nm photons.

As the ion beam from BRIF is delivered continuously, the DAQ is modified accordingly from that in Ref.~\cite{CLS-PKU}. Briefly, signals from the four PMTs are amplified and discriminated before being recorded by the TDC (TimeTagger4-2G time-to-digital converter, ChronoLogic). A digital-delay pulse generator (QC9528, Quantum Composers) provides the trigger signals with 1-kHz repetition rate, corresponding to a time window ($\Delta {\rm t}$) of 1~ms for the TDC. The DAQ program written in Python controls the reading of the events from TDC at each time interval of $n\Delta {\rm t}$. As introduced above, the HFS spectrum is measured by applying a scanning voltage ($\Delta U$) to the electrode before the CEC, with the laser frequency fixed. The scanning voltage between (at maximum) $-$2~kV and $+$2~kV is supplied by a Trek 623B DC amplifier with a gain of 1000. The DAQ program controls a USB device (USB-3106, Measurement Computing) to operate the input voltage. Special efforts were taken to ensure the correct synchronization between the measured photon counts and scanning voltage (see below in Sec.~\ref{sec:test}). The DAQ program also records the read-out voltage from the electrode via a digital multimeter (34470A, Keysight) together with a 1:1000 voltage divider (KV-10A, Ohm-labs). The HFS spectrum of the isotope of interest can be obtained by plotting the count rate of the fluorescent photons as a function of the scanning voltage.

\section{Test of neutralization and voltage tuning}\label{cectest}

To evaluate the general performance of the CEC and voltage scanning system, we preliminarily tested the neutralization and Doppler tuning using stable titanium and aluminium beams of 20-keV energy produced using an off-line laser ablation ion source at Peking University~\cite{CLS-PKU}. \mbox{Figure~\ref{Fig5}(a)} illustrates the schematic diagram for the voltage tuning and CEC heating in this test.

The CEC was loaded with chopped potassium chunks (Sigma Aldrich) and heated to $\sim$200$^{\circ}$C using a DC power supply. This power supply has an automatic feed-back control system, which allows the CEC temperature to be stabilized at a preset value. Before gradually heating the center, two ends of the CEC was kept at a temperature just above the melting point of K (63.2~$^{\circ}$C) by circulating the coolant (Ultra 350, LAUDA) using the circulator (ECO RE 1050S, LAUDA). This temperature was monitored by two PT100 thermometers, as indicated in Fig.~\ref{Fig5}(a). The influence of the CEC heating on the beamline vacuum was tested. For example, the vacuum changed only from 2.8$\times$10$^{-8}$ to 7$\times$10$^{-8}$~mbar when heating from room temperature to about 200$^{\circ}$C. This performance is essential for the future extension of the CLS setup to be compatible with high-sensitivity resonance ionization spectroscopy, which requires ultra-high vacuum to reduce the collisional ionization background~\cite{CRIS}. To avoid blocking the CEC apertures due to accumulation of solidified potassium, cooling the CEC ends needs to be handled carefully. The circulation of the 80$^{\circ}$C liquid can only be stopped when the readout temperature at the CEC center is about or below 80$^{\circ}$C, as shown in Fig.~\ref{Fig5}(b). For the current system, cooling down the CEC after operation takes roughly 3-4 hours.

\begin{figure}[b!]
\center
\includegraphics[width=0.5\textwidth]{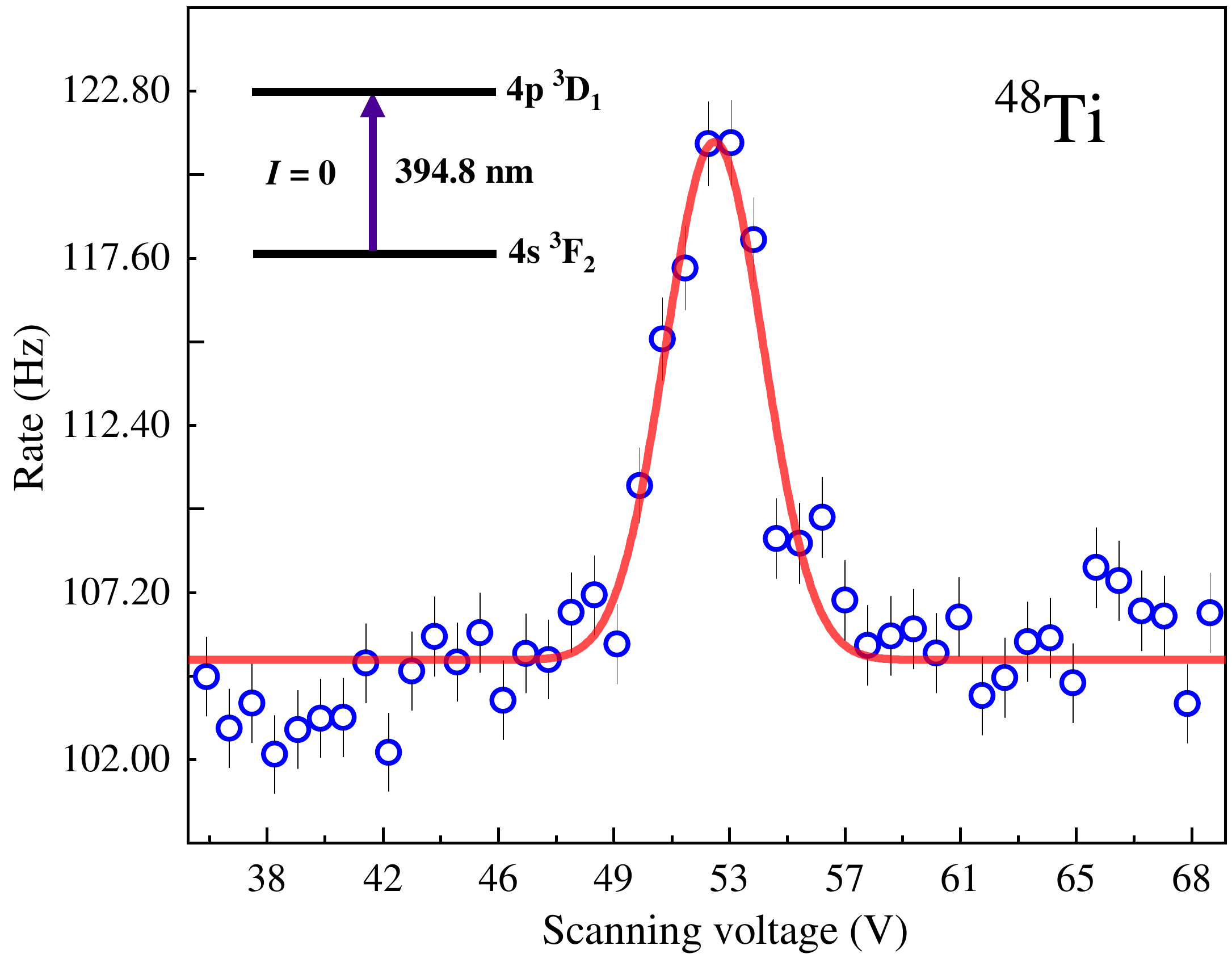}
\caption{Optical spectrum of $^{48}$Ti atom measured for $4s~^{3}\textrm{F}_{2}$ $\rightarrow$ 4$p$ $^{3}\textrm{D}_{1}$ transition (394.8~nm) by counting the emitted fluorescent photon as a function of scanned voltage applied to the CEC.}
\label{Fig6}
\end{figure}

As illustrated in Fig.~\ref{Fig5}(a), at $\sim$200$^{\circ}$C, the high density of potassium vapor should have been formed in the interaction region of the CEC. Thus, the incoming ions undergo charge exchange and are neutralized when passing through the potassium vapor. Using the electrostatic deflector mounted after the CEC, non-neutralized ions can be deflected away from the straight beam axis. The neutralization factor can then be measured by using the FC3 or the MageneTOF ion detector at the end of the CLS beamline. The neutralization efficiencies were measured to be approximately 15\% for Ti and 50\% for Al. The true temperature within the reservoir is probably lower than the temperature read by the thermocouple attached to its outer shell. 

\begin{figure}[t!]
\center
\includegraphics[width=0.45\textwidth]{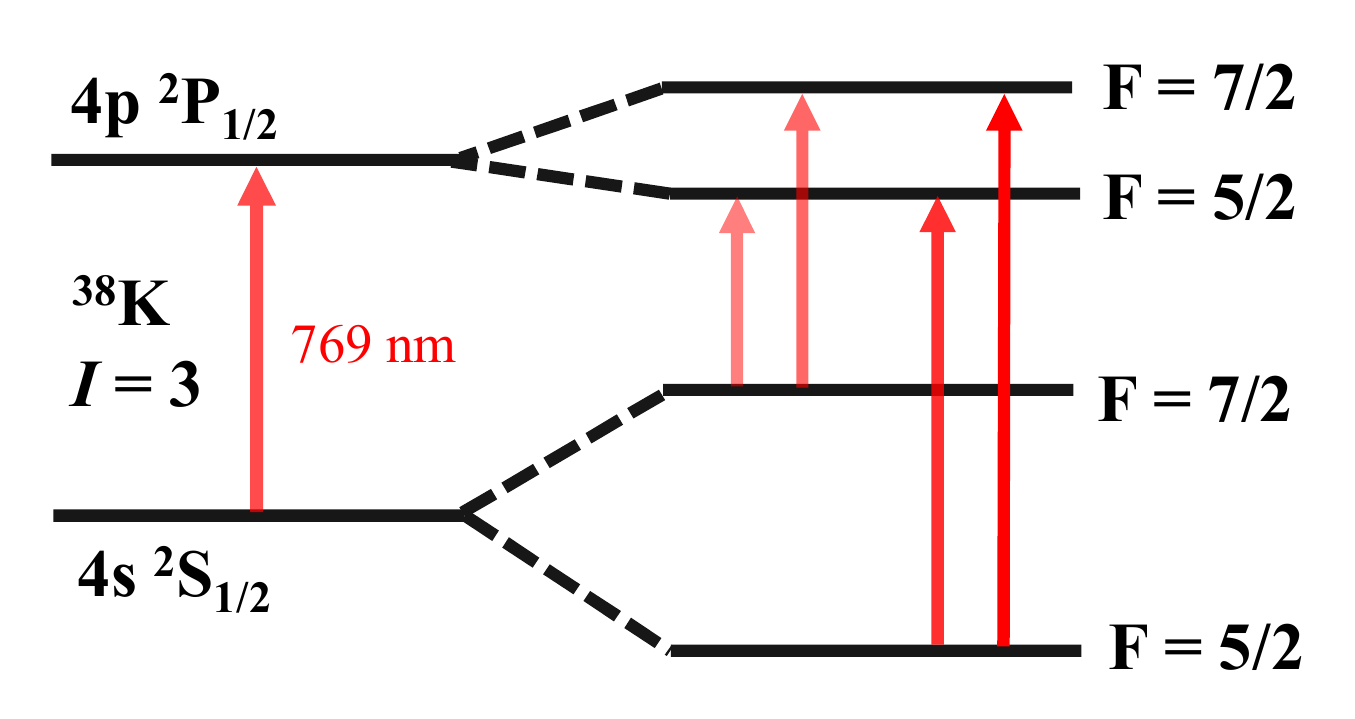}\\
\includegraphics[width=0.45\textwidth]{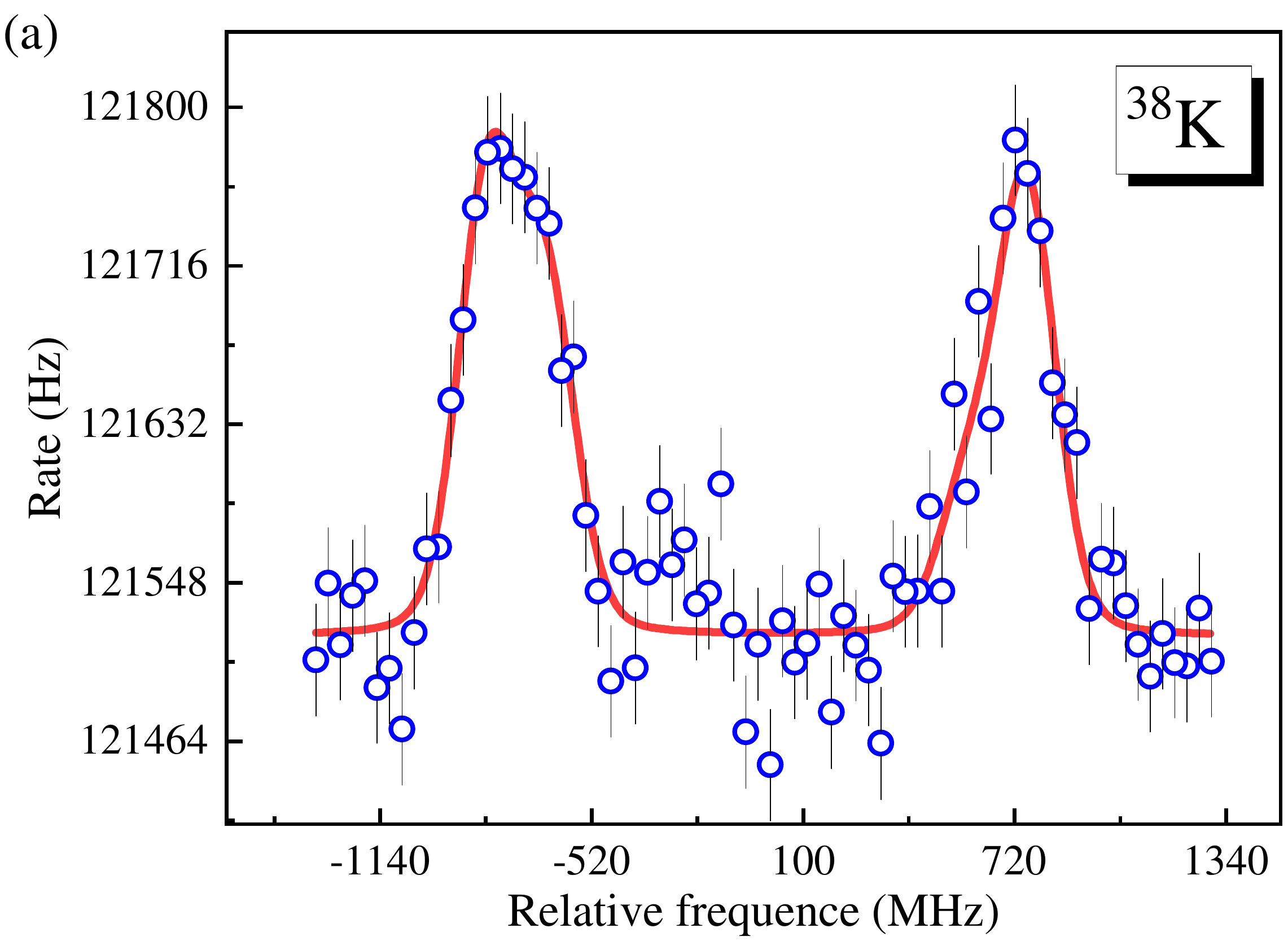}\\
\includegraphics[width=0.45\textwidth]{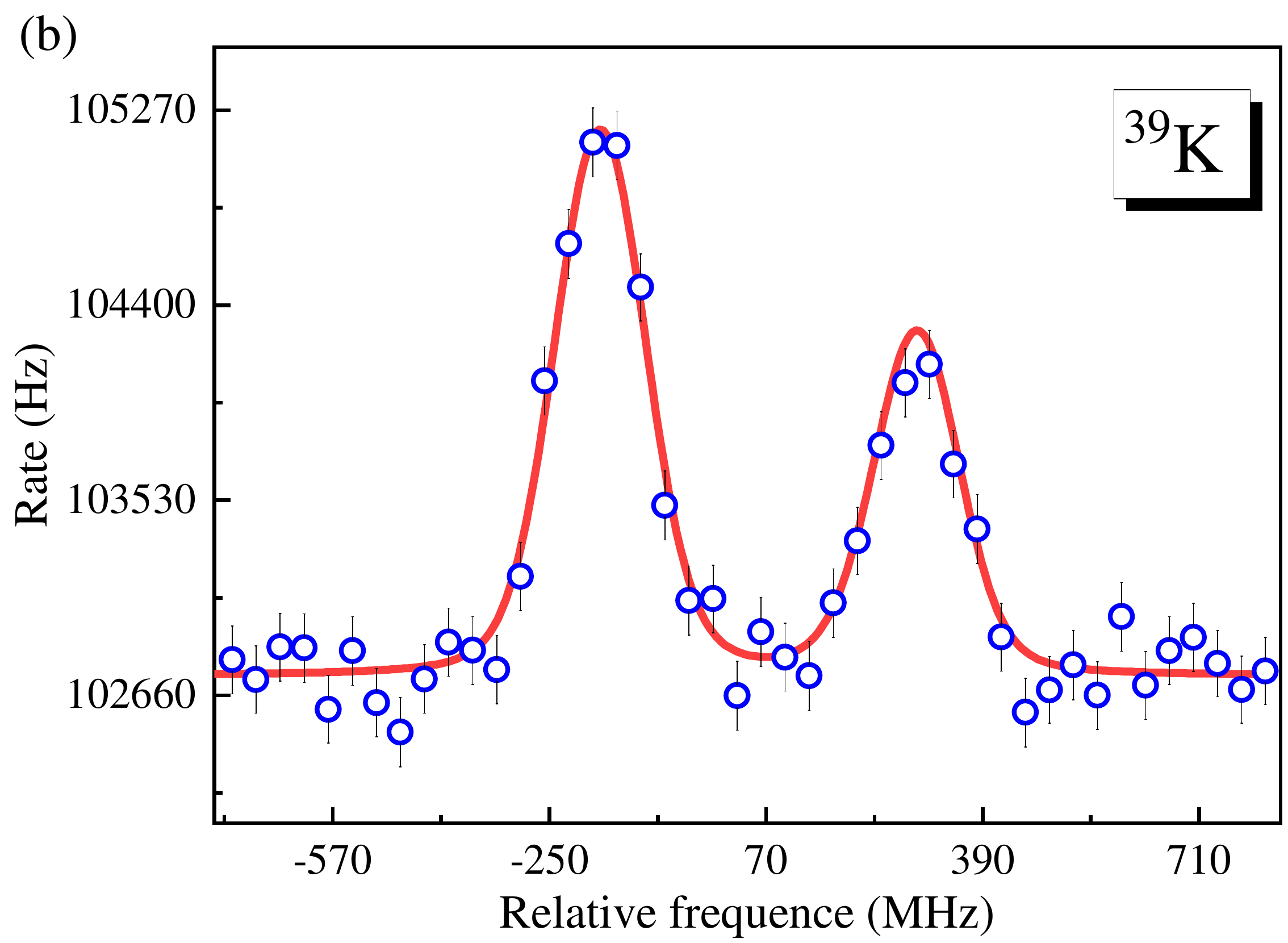}
\vspace{-2mm}
\caption{HFS spectra of $^{38,39}$K measured on the \mbox{$4s$ $^{2\!}S_{1/2}$} $\to$ \mbox{$4p$ $^{2\!}P_{1/2}$} (D1) atomic transition. Data are fitted with a Voigt line profile using SATLAS package~\cite{SATLAS}.}
\vspace{-2mm}
\label{Fig7}
\end{figure}

Tests of the voltage scanning were performed to probe the optical spectrum of 4$s$ $^{3}\textrm{F}_{2}$ $\rightarrow$ 4$p$ $^{3}\textrm{D}_{1}$ transition (394.8~nm) for $^{48}$Ti atom. The kinetic energy of the $^{48}$Ti ion beam was tuned by applying a scanning voltage to the CEC and the electrode, as shown in Fig.~\ref{Fig5}. With the laser frequency fixed to $\nu_0$, in the anti-collinear configuration, the accelerated (or decelerated) beam experiences a Doppler-shifted laser frequency $\nu$, which can be calculated as:
\begin{equation}
\nu=\nu_{0}\times\sqrt{\frac{1+\beta}{{1-\beta}}},\\
\label{eq:one}
\end{equation}
\begin{equation}
\beta=\sqrt{1-\frac{m^2c^4}{(E+mc^2)^2}},\\
\label{eq:two}
\end{equation}
where $E$ is the total kinetic energy and $m$ the mass of the atom. To resonantly excite the $^{48}$Ti atom, the wavenumber of the laser ($k=\nu/c$) is fixed to be 25293.9~cm$^{-1}$. Figure~\ref{Fig6} shows an example of the observed spectrum by counting the fluorescent photons emitted from the excited $^{48}$Ti atoms as a function of scanning voltage, fully demonstrating the performance of the CEC and voltage scanning.

\section{On-line commissioning experiment}\label{sec:test}

The CLS setup installed at BRIF (Fig.~\ref{Fig1}), was firstly commissioned with the stable ($^{39}$K) and unstable ($^{38}$K) isotopes, by measuring the HFS spectrum of their atomic \mbox{$4s$ $^{2\!}S_{1/2}$} $\to$ \mbox{$4p$ $^{2\!}P_{1/2}$} D1 transition (769~nm). As no off-line ion source is currently available at BRIF, both stable and unstable beams were produced by bombarding a CaO target with the 100-MeV proton beam from the cyclotron~\cite{NIMB-BRIF-2016}. The nuclear reaction products diffusing out of the thick target were ionized by a surface ion source heated to around 2000$^{\circ}$C, and then extracted and accelerated to 60~keV. Using the first-stage mass separator, the isotopes with mass number (e.g. $A=39$) were selected and sent to the CLS beamline. The total transmission efficiency of the ion beam through the entire CLS beamline was about 30-40\%, which fluctuated over time. The transmission efficiency was not fully optimized due to the limited beam time and that some ion optics elements require further modification to prevent charging on insulated ceramic surfaces. The ion beam (around 8-mm diameter) and laser beam ($\sim$5-mm diameter) were overlapped in an anti-collinear geometry, by using the two iris diaphragms with tunable diameter (Sec.~\ref{sec:optics}), two collimators with 8~mm diameter at two ends of the CEC (Sec.~\ref{sec:cec}) and two aperture arrays before and after the photon-detection region (Fig.~\ref{Fig1}).

The ion beam was firstly neutralized through the collisions with potassium vapor in the CEC. The neutralization efficiency for 60-keV potassium ion beam, estimated by using the MagneTOF ion detector and FC, was kept at about $\sim$40\% during the experimental measurement. With the laser wavenumber fixed to 12691.35 cm$^{-1}$, HFS spectra of neutral $^{38,39}$K were obtained by measuring the laser-induced fluorescence as a function of the scanning voltage applied to the CEC.

Figure~\ref{Fig7} presents the HFS spectra of $^{38,39}$K measured with only two PMTs in the second row downstream of the CEC. The other two PMTs in the first row did not function correctly during the experiment. The expected four HFS resonances (within two pairs) for $^{38,39}$K isotopes are reasonably reproduced but the narrowly-spaced pairs of peaks resulting from the magnetic dipole splitting of the upper state are not well resolved. This is due to the large energy spread of the ion beam. The contribution of Gaussian linewidth is estimated to be $\sim$ 125~MHz by fitting the HFS spectra in Fig.~\ref{Fig7} using a Voigt profile (a convolution of the Gaussian function and the Lorentzian function). This corresponds to an energy spread of \mbox{$\sim$21~eV}, estimated based on $\delta \nu=\nu_{0}\times(\delta E)/(\sqrt{2Emc^2})$.

It is worth noting that the rate (Hz) is used in Fig.~\ref{Fig7} for the $y$ axis. This rate is defined as the number of counts divided by the counting time for each voltage step (corresponding to each frequency). The step size is normally a few volts and preset by the DAQ program. At each voltage step (normally sustained for a few tens of ms), the DAQ will execute repeatedly the \lq read' function to read the events from TDC for, e.g. 20 times, then switch to the next voltage. In this way, the correct synchronization between the photon counts at each voltage is ensured. Since the reading time at each voltage step is not necessarily identical, the counts/time (rate) for each voltage setting is meaningful here and is used instead of the total number of counts. Figure~\ref{Fig6} is plotted in the same way, but having a much better signal-to-background ratio owing to the bunched ion beam produced off-line at Peking University. This spectroscopic method has already been adopted in the literature for measurements using similar TDC device~\cite{IGISOL,CRIS}.

\begin{table*}[!t]
\centering
\setlength{\abovecaptionskip}{0pt}
\setlength{\belowcaptionskip}{10pt}
\caption{Magnetic dipole hyperfine constants of $^{38,39}$K isotopes for the two atomic states, and isotope shift (difference in the center-of-gravity frequencies) between these two isotopes. These results (given in MHz) are compared to the literature values~\cite{K-radii2021,K-AB-2019,PRC-K-2014}.}
\renewcommand*{\arraystretch}{1.2}
\begin{tabular}{c |c c |c c| c c}
\hline
Isotope &$A_{\rm l}$ &~$A_{\rm l}$ \cite{K-AB-2019}/\cite{PRC-K-2014} &$A_{\rm u}$ &$A_{\rm u}$ \cite{K-AB-2019}/\cite{PRC-K-2014} & $\delta \nu^{39,A}$ &$\delta \nu^{39,A}$\cite{K-radii2021}\\
\hline
$^{38}$K & 404(6) & 404.2(5)/404.3(3) & 49(4) & 49.0(6)/48.9(2) & -125(9) & -126.1(19) \\
$^{39}$K & 230(2) & 231.1(3)/231.0(3) & 29(3) & 27.1(6)/27.8(2) & - & - \\
\hline
\end{tabular}
\label{tab1}
\end{table*}

To extract the atomic parameters, the HFS spectrum is fitted using a $\chi^2$-minimization Python routine in SATLAS~\cite{SATLAS}. The frequency of each resonance peak can be described by the function~\cite{JPG2010}:
\begin{equation}
\nu=\nu_{0}+\alpha_{\rm u}A_{\rm u}-\alpha_{\rm l}A_{\rm l}\\
\label{eq:one}
\end{equation}
where $\nu_{0}$ is centroid frequency, $A_{\rm u}$ and $A_{\rm l}$ are the magnetic dipole hyperfine constants for the excited state ($^{2\!}P_{1/2}$) and ground state ($^{2\!}S_{1/2}$), respectively, $\alpha =\frac{1}{2} \lbrack F(F+1)-I(I+1)-J(J+1)\rbrack$ with $J$ the total angular momentum for each atomic state. In this analysis, the ratios of $A_{\rm u}$ and $A_{\rm l}$ from the literature~\cite{K-AB-2019,PRC-K-2014} are used as the constraint but allowed to vary within the uncertainty, due to the relatively wide line-width of the observed HFS spectra. The extracted magnetic dipole hyperfine constants ($A_{\rm l}$ and $A_{\rm u}$) of $^{38,39}$K for the two atomic states, and isotope shift (the difference in the centroid frequencies) of the two isotopes, are consistent with literature values within the errors ~\cite{K-radii2021,K-AB-2019,PRC-K-2014}, as summarized in Table~\ref{tab1}.

From the measured HFS spectrum of $^{39}$K in Fig.~\ref{Fig7}(b), we can briefly estimate the detection efficiency measured by two fully functioning PMTs. The count rate above background for the most intense resonance is about 2500~${\rm s}^{-1}$. Taking the estimated neutral beam current of $\sim$55~pA and assuming 100\% purity of the beam, the combined excitation and detection efficiency is 1:130000. We expect around a factor of 2 increase in overall efficiency (1:65000) if all four PMTs were functional for the experiment. This is of the same order of magnitude as that of similar CLS setups reported in literature~\cite{BECOLA}.

\section{Summary and outlook}

We have developed a collinear laser spectroscopy system and installed it at BRIF. By using stable ($^{39}$K) and unstable ($^{38}$K) ion beams produced at BRIF, this CLS system was commissioned successfully by probing the hyperfine structure of the atomic \mbox{$4s$ $^{2\!}S_{1/2}$} $\to$ \mbox{$4p$ $^{2\!}P_{1/2}$} transition (769~nm) of these isotopes. The estimated overall detection efficiency is of the same order of magnitude as those reported in the literature for the same type of CLS setup. The extracted magnetic dipole hyperfine constants and the isotope shift of $^{38,39}$K are in good agreement with the literature values, demonstrating the overall performance of the present CLS setup. 

Although substantial progress has been achieved in this work, problems, such as the relative large spectra linewidth ($\sim$150~MHz), were noticed. This is mostly attributed to the energy spread of the ion beam delivered by the BRIF facility, which is estimated to be approximately 21~eV for the ion beam at 60~keV. Further development and installation of a RFQ cooler-buncher~\cite{ISCOOL} are now planned, which would provide bunched ion beams with smaller energy spreads of a few electron volts, allowing a large improvement in resolution and sensitivity of the CLS setup.

This successful operation of the CLS setup has opened new opportunities for laser spectroscopy measurement of unstable isotopes at BRIF and other upcoming radioactive ion beam facilities in China. For example, high-precision charge radii measurement of neutron-deficient Al and Na isotopes are expected at BRIF after the implementation of the RFQ cooler buncher. With the on-going development of the UC$_{x}$ target at BRIF and the construction of the HIAF RIB facility, high-resolution laser spectroscopy measurement of unstable isotopes in the heavier mass region (e.g. the lead region) can be anticipated in the near future.\\

\section*{Acknowledgments}

We acknowledge the support of the BRIF collaboration and technical teams. The authors thank the members from CRIS and COLLAPS collaboration for the useful discussion and support on the design and development of this collinear laser spectroscopy system. This work was supported by National Key R\&D Program of China (No. 2018YFA0404403),  National Natural Science Foundation of China (Nos. 12027809, U1967201, 11875073, 11875074, 11961141003 and 12125509), China National Nuclear Corporation (No. FA18000201), and the State Key Laboratory of Nuclear Physics and Technology, Peking University (No. NPT2019ZZ02). 

\bibliographystyle{elsarticle-num}
\bibliography{CLS-BRIF}

\end{document}